# Optimal orientation of striped states in the quantum Hall system against external modulations

**T Aoyama, K Ishikawa, and N Maeda**

Department of Physics, Hokkaido University, Sapporo 060-0810, Japan

**Abstract.** The physical properties of striped Hall state in the presence of an external unidirectional periodic potential is investigated. The study of searching the optimal orientation of the state is reported for the state at filling factor $\nu = 2 + 1/2$ and for the state at higher Landau levels at $l + 1/2$ where $l = 3, 4, 5, 6, 7, 8$. It is predicted that two phases of the orientation, parallel phase and perpendicular phase, exist and a phase transition from one to another phase occurs.

In two-dimensional (2D) electron systems under a strong magnetic field, highly anisotropic states have been observed at high Landau Levels (LL)[1, 2, 3, 4]. The resistance in one direction is very different from that of another direction. This state seems to agree with striped state which was predicted in a mean field theory[5, 6]. The Hamiltonian of 2D electron system in a perpendicular magnetic field is invariant under translation and rotation, but these symmetries are broken spontaneously in the striped state which chooses a particular direction. The mechanism of choosing the orientation is unknown. The origin of the anisotropic orientation is naively considered as surface morphology of semiconductors, crystal structure and defects. The small perturbation such as a crystallographic axis will be sufficient to determine the orientation. The anisotropic transport in unidirectional lateral super lattices [7] and the interchange of anisotropy by varying the 2D electron density[8] have been observed recently. In this work, we demonstrate that an external modulation determines the orientation of anisotropy[9] and compare our results with experiments.

In the von Neumann lattice (vNL) formalism[10, 11], which uses complete sets of coherent state of guiding center variables, we transform the 2D electron system to the 2D lattice electron system. The momentum is a useful parameter in this representation. The striped states have a Fermi surface and physical property of striped states is understandable from the shape of the Fermi surface.

Let us consider a 2D electron system in a perpendicular uniform magnetic field $B$. The total Hamiltonian $H$ of the system is written as $H = H_0 + H_1$,

$$H_0 = \int \psi^\dagger(\mathbf{r}) \frac{(\hat{\mathbf{p}} + e\mathbf{A})^2}{2m} \psi(\mathbf{r}) d^2 r,$$
$$H_1 = \frac{1}{2} \int \rho(\mathbf{r}) V(\mathbf{r} - \mathbf{r}') \rho(\mathbf{r}') d^2 r d^2 r', \tag{1}$$

where $\hat{p}_\alpha = -i\hbar\partial_\alpha$, $\partial_x A_y - \partial_y A_x = B$, $V(\mathbf{r}) = q^2/r$, $q^2 = e^2/4\pi\epsilon$ ($\epsilon$ is a dielectric constant). $\psi(\mathbf{r})$ is the electron field, and $\rho(\mathbf{r}) = \psi^\dagger(\mathbf{r})\psi(\mathbf{r})$. $H_0$ is the free Hamiltonian, which is quenched in the LL. $H_1$ is the Coulomb interaction. We ignore the spin degree of freedom. The electron field is expanded by the momentum state $|f_l \otimes \beta_{\mathbf{p}}\rangle$ in vNL formalism[10, 11] as

$$\psi(\mathbf{r}) = \sum_{l=0}^{\infty} \int_{\mathrm{BZ}} \frac{d^2 p}{(2\pi)^2} b_l(\mathbf{p}) \langle \mathbf{r} | f_l \otimes \beta_{\mathbf{p}} \rangle, \tag{2}$$

where $b_l(\mathbf{p})$ is the anticommuting annihilation operator with a LL index $l$ and a 2D lattice momentum $\mathbf{p}$ defined in the Brillouin zone (BZ) $|p_\alpha| \leq \pi$. The annihilation operator $b_l(\mathbf{p})$ obeys a twisted periodic boundary condition $b_l(\mathbf{p} - 2\pi\mathbf{N}) = e^{-i\pi(N_x+N_y)+iN_yp_x}b_l(\mathbf{p})$, where $N_x$, $N_y$ are integers. The momentum state is Fourier transform of the Wannier basis of vNL which localized at $\mathbf{r} = a(r_sm, n/r_s)$, where $n$, $m$ are integers. Here $a = \sqrt{2\pi\hbar/eB}$ and $r_s$ is an asymmetric parameter. We consider only the $l$-th LL state and ignore the LL mixing. Hence, $H_0$ turns out to be constant and we omit the free Hamiltonian. Fourier transformed density operator $\tilde{\rho}$ in the $l$-th LL is written in vNL formalism as

$$\tilde{\rho}(\mathbf{k}) = \int_{\mathrm{BZ}} \frac{d^2p}{(2\pi)^2} b_l^\dagger(\mathbf{p}) b_l(\mathbf{p} - a\hat{\mathbf{k}}) f_l(k) e^{-i\frac{a}{4\pi}\hat{k}_x(2p_y - a\hat{k}_y)}, \tag{3}$$

where $\hat{\mathbf{k}} = (r_sk_x, k_y/r_s)$ and $f_l(k) = L_l(\frac{a^2k^2}{4\pi})e^{-\frac{a^2k^2}{8\pi}}$. Here $L_l$ is the Laguerre polynomial. The filling factor is fixed at $l + 1/2$ in the following calculation.

In the HF approximation, the Coulomb interaction term is reduced to the kinetic term by using a mean field $U(p) \equiv \langle\Psi_1|b_l^\dagger(\mathbf{p})b_l(\mathbf{p})|\Psi_1\rangle$

$$\begin{aligned} H_1 &= \frac{1}{2}\int \rho(\mathbf{r})V(\mathbf{r} - \mathbf{r}')\rho(\mathbf{r}')d^2rd^2r' \\ &\overset{\mathrm{HF}}{\simeq} \frac{\nu^*}{N}\int_{\mathrm{BZ}} \frac{d^2p}{(2\pi)^2}\frac{d^2p'}{(2\pi)^2} v_l^{\mathrm{HF}}(\mathbf{p} - \mathbf{p}')U(\mathbf{p}')b_l^\dagger(\mathbf{p})b_l(\mathbf{p}) \\ &\quad - \frac{1}{2}\left(\frac{\nu^*}{N}\right)^2 \int_{\mathrm{BZ}} \frac{d^2p}{(2\pi)^2}\frac{d^2p'}{(2\pi)^2} U(\mathbf{p})v_l^{\mathrm{HF}}(\mathbf{p} - \mathbf{p}')U(\mathbf{p}') \\ &= \int_{\mathrm{BZ}} \frac{d^2p}{(2\pi)^2}\epsilon_l^{\mathrm{HF}}(\mathbf{p})b_l^\dagger(\mathbf{p})b_l(\mathbf{p}) + \mathrm{const}, \end{aligned} \tag{4}$$

where $N$ is a number of electrons. $|\Psi_1\rangle$ is a many particle state which is determined by a self consistent condition. We define

$$v_l^{\mathrm{HF}}(\mathbf{p} - \mathbf{p}') = \sum_n \left( V_l\left(\frac{2\pi\tilde{\mathbf{n}}}{a}\right) e^{in_y(p-p')_x - in_x(p-p')_y} - V_l\left(\frac{2\pi\tilde{\mathbf{n}} + \tilde{\mathbf{p}}' - \tilde{\mathbf{p}}}{a}\right)\right), \tag{5}$$

where $\tilde{\mathbf{n}} = (n_x/r_s, n_yr_s)$ and $V_l(k) \equiv \frac{2\pi}{k}(f_l(k))^2$. The one-particle spectrum generated by the Coulomb interaction is defined by

$$\epsilon_l^{\mathrm{HF}}(\mathbf{p}) = \frac{\nu^*}{N}\int_{\mathrm{BZ}} \frac{d^2p'}{(2\pi)^2} v_l^{\mathrm{HF}}(\mathbf{p} - \mathbf{p}')U(\mathbf{p}'). \tag{6}$$

The explicit form of the mean field is $U(\mathbf{p}) = (2\pi)^2\frac{N}{\nu^*}\theta(\mu - \epsilon^{\mathrm{HF}}(\mathbf{p}))$ which is derived by using the anticommutation relation of the electron operator $\{b_l(\mathbf{p}'), b_l^\dagger(\mathbf{p})\} = (2\pi)^2\sum_n \delta^{(2)}(\mathbf{p}' - \mathbf{p} - 2\pi\mathbf{n})e^{i\pi(n_x+n_y)-in_yp_x}$. $\theta(t)$ is the step function. From Eq. (6), the self consistent condition is given by

$$\epsilon_l^{\mathrm{HF}}(\mathbf{p}) = \int_{\mathrm{BZ}} d^2p' v_l^{\mathrm{HF}}(\mathbf{p} - \mathbf{p}')\theta(\mu - \epsilon_l^{\mathrm{HF}}(\mathbf{p})). \tag{7}$$

In the following, we use the striped state $|\Psi_1\rangle$ which is uniform in the $y$-direction. This is given as

$$|\Psi_1\rangle = N_1 \prod_{|p_x|\leq\pi, |p_y|\leq\pi/2} b_l^\dagger(\mathbf{p})|0\rangle, \tag{8}$$

where $|0\rangle$ is the vacuum state for $b_l$ and $N_1$ is a normalization factor. This striped state is the most stable compressible state under the HF approximation at the half-filled higher

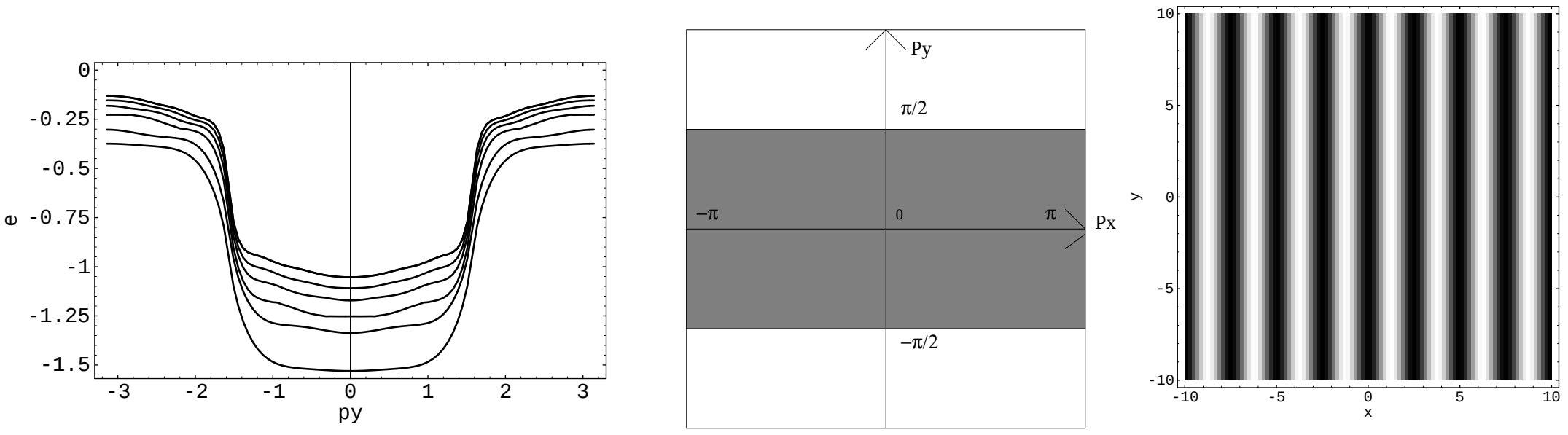

**Figure 1.** The left figure is the spectrum $\epsilon_l^{\rm HF}(p_y)$. The six curves correspond to six different LL's: $l = 2, \cdots, 7$ from bottom to top. The Fermi Sea is drawn in one BZ at half-filled LL's in the center figure. The charge density distribution is given in the right figure. The Fermi Sea is orthogonal to the stripe in coordinate space. The unit of $x, y$ and spectrum is $a$ and $q^2/a$, respectively.

LL[10, 12]. The Fermi surface is parallel to the $p_x$ axis (see Fig. 1). The density profile $\langle \Psi_1 | \rho(\mathbf{r}) | \Psi_1 \rangle$ is uniform in $y$ direction and periodic in $x$ direction with a period $ar_s$ (see Fig.1 ). The orthogonality of the Fermi surface in the momentum space and the density profile in the coordinate space plays important roles and is *reminiscent of the Hall effect.* The one-particle spectrum is given by $\epsilon_l^{\rm HF}(p_y) = \epsilon_l^{\rm H}(p_y) + \epsilon_l^{\rm F}(p_y)$,

$$\epsilon_l^{\rm H}(p_y) = \frac{2r_s q^2}{a\pi} \sum_{n=odd} f_l\left(\frac{2\pi n}{ar_s}\right)^2 \frac{\cos(np_y)}{n^2}, \tag{9}$$

$$\epsilon_l^{\rm F}(p_y) = -\frac{r_s q^2}{a} \sum_{n=-\infty}^{\infty} \int_{-\frac{\pi}{2}+p_y+2\pi n}^{\frac{\pi}{2}+p_y+2\pi n} dk_y \int_{-\infty}^{\infty} \frac{dk_x}{2\pi} h_l(\mathbf{k}), \tag{10}$$

where $h_l(\mathbf{k}) = f_l(\sqrt{k_x^2 + (k_y r_s/a)^2})^2 / \sqrt{k_x^2 + (k_y r_s/a)^2}$. $\epsilon_l^{\rm HF}$ depends on only $p_y$, and satisfies with the self consistent condition Eq. (7). The one-particle spectrum has an energy gap in $p_x$ direction and is gapless in $p_y$ direction. The HF energy per particle is calculated as $E_l^{\rm HF}(r_s) = \langle \Psi_1 | H_1 | \Psi_1 \rangle / N$. The optimal value $r_s = r_s^{\rm min}$ is determined so as to minimize $E_l^{\rm HF}(r_s)$ (see Fig. 2)[10]. The optimal values of $r_s^{\rm min}$ and corresponding energy $E_l^{\rm HF}$ are given in Table. 1.

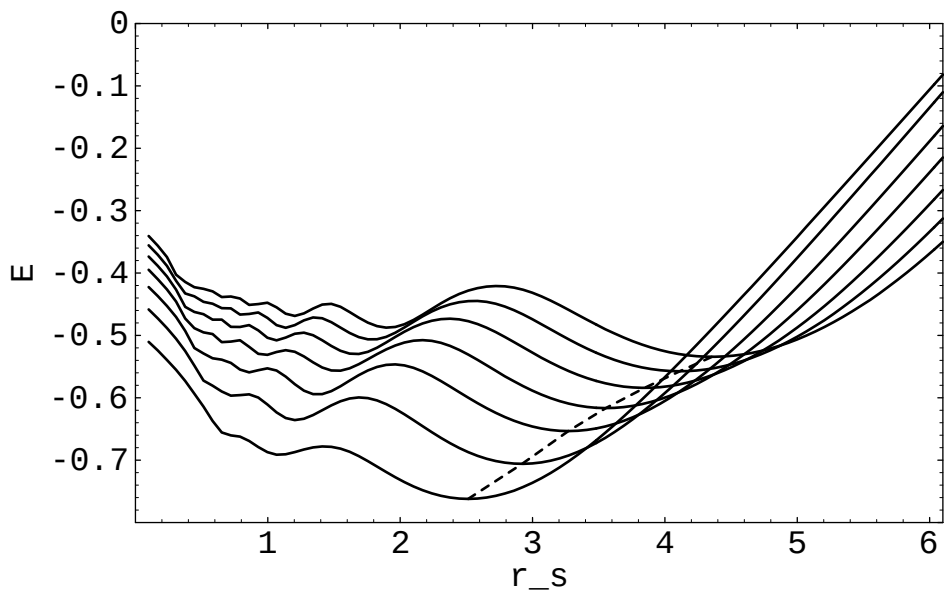

**Figure 2**. The HF energy per particle $E_l^{\rm HF}(r_s)$. The seven curves correspond to seven different LL's: $l = 2, \cdots, 8$ from bottom to top. The dashed line represents the minimum value of $E_l^{\rm HF}(r_s)$ for each LL's. The unit of $r_s$ and $E_l^{\rm HF}$ is $a$ and $q^2/a$, respectively.

| $l$ | $r_s^{\rm min}$ | $E_l^{\rm HF}(r_s^{\rm min})$ |
|---|---|---|
| 2 | 2.474 | -0.771 |
| 3 | 2.921 | -0.706 |
| 4 | 3.271 | -0.653 |
| 5 | 3.554 | -0.617 |
| 6 | 3.847 | -0.584 |
| 7 | 4.113 | -0.557 |
| 8 | 4.356 | -0.534 |

**Table 1**. The optimal value $r_s = r_s^{\rm min}$ and corresponding energy $E_l^{\rm HF}$. The unit of $r_s$ and $E_l^{\rm HF}$ is $a$ and $q^2/a$, respectively.

Let us consider the effect of an external modulation. The total Hamiltonian $H$ of the system is written as $H = H_0 + H_1 + H_2$, where $H_0 + H_1$ is given in Eq. (1) and $H_2$ is an external modulation

$$H_2 = g \int \rho(\mathbf{r}) \cos(\mathbf{K} \cdot \mathbf{r}) d^2 r. \tag{11}$$

As mentioned before, $H_0$ is quenched in the LL. We substitute Eq. (3) into $H_1$ and $H_2$ and find the ground state in two perturbative approaches. In the first approach, perturbative expansions with respect to $g$ in $H_2$, which describe the Coulomb dominant regime, are applied. In the second approach, perturbative expansions with respect to $q^2/a$ in $H_1$, which describe the external modulation dominant regime, are applied.

**External modulation as a perturbation** : We obtain the ground state of $H_1$ in the HF approximation first. Using the HF ground state, we treat the external modulation $H_2$ as a perturbation. This approximation is relevant in the Coulomb dominant regime $g \ll q^2/a$. The perturbation energy per particle in the first order $\Delta E_l^{(1)} = \langle \Psi_1 | H_2 | \Psi_1 \rangle / N$ is written as

$$\Delta E_l^{(1)} = \frac{g}{2N} \langle \Psi_1 | (\tilde{\rho}(\mathbf{K}) + \tilde{\rho}(-\mathbf{K})) | \Psi_1 \rangle. \tag{12}$$

Here, we use $H_2 = \frac{g}{2}(\tilde{\rho}(\mathbf{K}) + \tilde{\rho}(-\mathbf{K}))$ which is written by the Fourier transformed density. The operator $\tilde{\rho}(\mathbf{K})$ moves an electron in the Fermi sea by $a\hat{\mathbf{K}}$ in the momentum space, and thus many particle state $\tilde{\rho}(\pm\mathbf{K})|\Psi_1\rangle$ is orthogonal to the striped state $|\Psi_1\rangle$. Therefore, except for the case that $a\hat{K}_y$ coincides with a multiple of $2\pi$, $\Delta E_l^{(1)}$ vanishes. We consider only the range $|a\hat{K}_y| < \pi$ which is sufficient to compare our results with experiments. The perturbation energy per particle in the second order $\Delta E_l^{(2)}$ is written as

$$\Delta E_l^{(2)}(g, \theta, K) = \int_{\frac{\pi}{2} - a\hat{K}_y}^{\frac{\pi}{2}} \frac{dp_y}{2\pi} \frac{-g^2 f_l(K)^2}{\epsilon_l^{\mathrm{HF}}(p_y + a\hat{K}_y) - \epsilon_l^{\mathrm{HF}}(p_y)}, \tag{13}$$

where $\hat{K}_y = K \sin\theta / r_s^{\min}$, $\theta$ is an angle between the stripe direction and external modulation. We obtain the total energy per particle in the Coulomb dominant regime at the second order perturbation as

$$E_l^{\mathrm{Coul}}(g, K, \theta) = E_l^{\mathrm{HF}}(r_s^{\min}) + \Delta E_l^{(2)}(g, K, \theta). \tag{14}$$

For example, the $\theta$ dependence of $\Delta E_l^{(2)}$ is shown in Fig. 3 at the half-filled third LL. As seen in Fig. 3, the energy is always minimum at $\theta = \pi/2$, that is, the optimal orientation of the striped state is *orthogonal to the external modulation*. We call this phase the orthogonal phase. Note that $\Delta E_l^{(2)}$ vanishes and $\theta$ dependence disappears when $K$ equals the zeros of $f_l(K)$. In this case, the external modulation loses control of the stripe direction.

**Coulomb interaction as a perturbation** : In the external modulation dominant regime $g \gg q^2/a$, it is expected that the external modulation makes the striped state to be commensurate to the external modulation. For simplicity, the external modulation is parallel to $y$ axis, that is $H_2 = g \int \rho(x) \cos(K_x x) dx$. Using the twisted periodic boundary condition $b_l(\mathbf{p})$, we diagonalize $H_2$ by taking $r_s = 2\pi/aK$ $(K_x \equiv K)$, which means that the stripe state is commensurate with the external modulation. The period of the stripe states eqauls to the wave length $2\pi/aK$ of the external modulation, not to the optimal value $r_s^{\min}$ determined to minimize $E_l^{\mathrm{HF}}(r_s)$. We get, then, the spectrum of $H_2$ depending only on $p_y$ momentum. The spectrum satisfies the self consistent condition because $\epsilon_l^{\mathrm{HF}}$ depends on only $p_y$ at the half-filling for the highest LL. Using this vNL basis, we treat the Coulomb interaction $H_1$ as a perturbation. Then the state $|\Psi_1\rangle$ is the ground state of $H_2$. The external modulation term reads

$$H_2 = -|g f_l(K)| \int_{\mathrm{BZ}} \frac{d^2 p}{(2\pi)^2} b_l^\dagger(\mathbf{p}) b_l(\mathbf{p}) \cos p_y. \tag{15}$$

The ground state energy per particle of $H_2$ is obtained as $E_2(g) = -\frac{2}{\pi}|gf_l(K)|$. The perturbation energy per particle is calculated as $E_l^{\rm HF}(\rm r_s) = \langle\Psi_1|\rm H_1|\Psi_1\rangle/N$ with $r_s = 2\pi/aK$. Hence, the total energy per particle in the external modulation dominant regime is given by

$$E_l^{\rm ext}(g,K) = E_l^{\rm HF}(2\pi/aK) - \frac{2}{\pi}\left|gf_l(K)\right|. \tag{16}$$

In this state, the stripe direction is *parallel to the external modulation*, which we call the parallel phase. We compare the total energies of striped states obtained in the orthogonal and the parallel phase. As a typical example, $E_l^{\rm Coul}(g,K,\pi/2)$ and $E_l^{\rm ext}(g,K)$ for $aK = 2$ is plotted in Fig. 4 at the half-filled third LL. The bold line represents the lower energy state. As seen in Fig. 4, the orthogonal phase has lower energy and is realized at small $g$. The parallel phase has lower energy and is realized at large $g$. The phase boundary is calculated by solving the equation in $g$, $E_l^{\rm ext}(g,K) = E_l^{\rm Coul}(g,K,\pi/2)$ for various value of $aK$. The phase diagram in $g$-$K$ plane is shown in Fig. 5, where orthogonal and parallel phases are indicated by I and II respectively. The dashed lines correspond to the zeros of $f_l(K)$, at which the stripe direction is undetermined. When the period of the external modulation coincides with the optimal period of the stripe, the phase boundary touches the $K$ axis. The wave number value on the phase boundary at $g = 0$ is inverse proportional to $l$, due to the optimal period of the stripe is proportional to $l$.

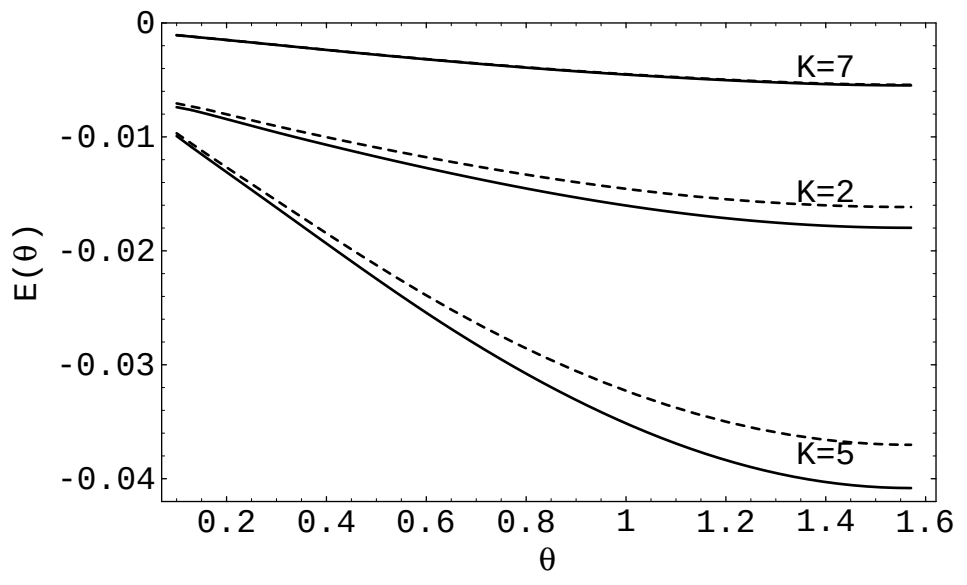

**Figure 3.** The $\theta$-dependence of $\Delta E_l^{(2)}(g,K,\theta)$ for $aK = 2, 5, 7$. The solid lines and dashed lines stand for the HF calculation and RPA approximation, respectively. The unit of energy is $g^2a/q^2$

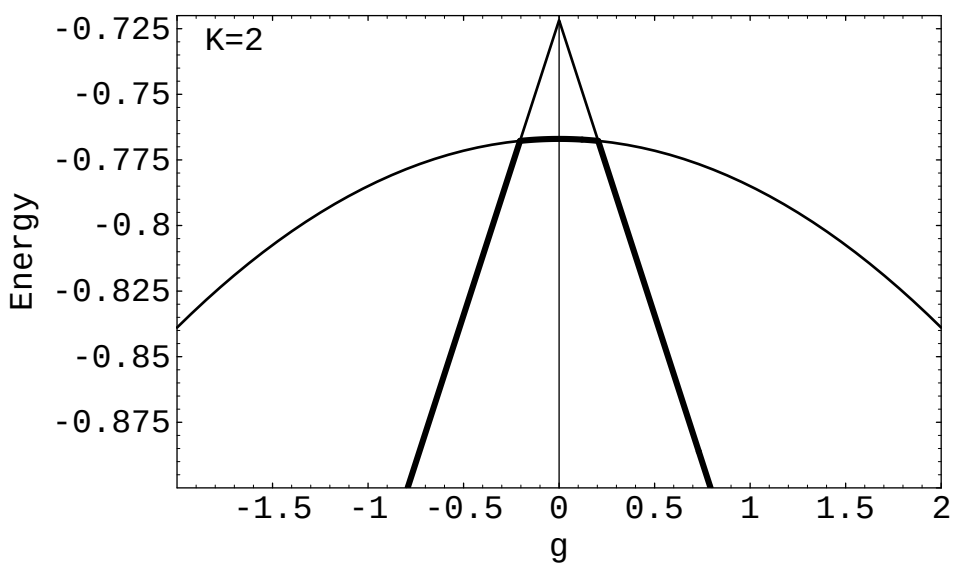

**Figure 4.** The total energy of the striped state obtained in (I) and (II) for $aK = 2$. The unit of energy and $g$ is $q^2/a$. The straight line stands for $E_l^{\rm ext}(g, 2/a)$ and the parabolic line stands for $E_l^{\rm Coul}(g, 2/a, \pi/2)$. The lower energy state is represented by the bold line.

The direct verification of our results is made by observing a transition between the two phases by tuning the extenal modulation. The necessary wave length of the modulation for the verification is on the order of $2a$, which is about 100nm at $B = 2\rm T$. The unidirectional lateral superlattice with a period 92nm has been achieved on top of the 2D electron system[7, 13]. The experiment shows that the magnetoresistance orthogonal to the external modulation has a shallow and broad dent between two peaks around $\nu = 9/2$. The magnetoresistance parallel to the external modulation does not have the same structure around $\nu = 9/2$. Anisotropy observed in this experiment is small due to the low mobility compared with experiments of striped states[1, 2, 3, 4]. The strength of the modulation is estimated as $g = 0.015\rm meV$. The parameters $(g,K) = (0.006q^2/a, 3.097/a)$ correspond to this experimental setting, and are shown as $X$ in Fig. 6. $X$ belongs to the orthogonal phase. In this phase, the one-particle dispersion has no energy gap in the orthogonal direction and has an energy gap in the parallel direction to the external modulation. Hence, the magnetoresistance orthogonal to the external modulation is strongly modified by the injected electric current compared with the parallel magnetoresistance. This is consistent with the experiment. With a slightly larger period 115nm, orthogonal magnetoresistance around $\nu = 9/2$ is structureless. In this case, the corresponding parameters $(g, K) = (0.014q^2/a, 2.478/a)$ are shown as $Y$ in Fig. 6[7, 13]. $Y$ belongs to the parallel phase. Hence, there is an energy gap in orthogonal direction to the external modulation and the magnetoresistance in this direction is not modified strongly by the injected electric current. This is also consistent with the experiment.

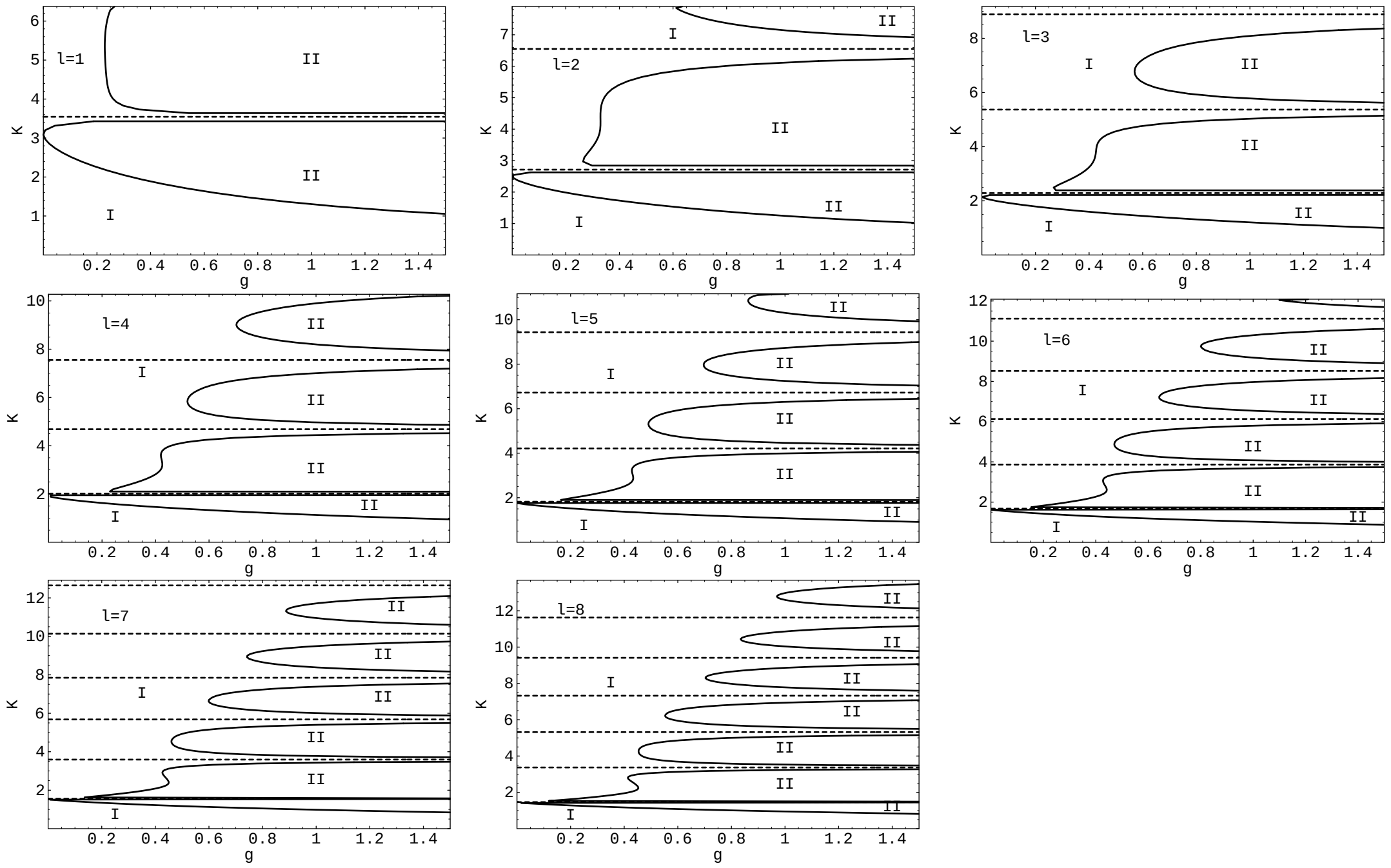

**Figure 5**. The phase diagram of the striped state at the half-filled $l+1$-th LL. The unit of $g$ is $q^2/a$ and the unit of $K$ is $1/a$. The region denoted by I corresponds to the orthogonal phase. The region denoted by II corresponds to the parallel phase. The dashed lines represent the zeros of $f_l(K)$, at which the stripe direction is undetermined.

Recently, the experiment tuning the external modulation by changing the 2D electron density, has been done at the half-filled $l+1$-th LL for $l = 1, 2, 3, 4, 5, 6$[7]. The magnetoresistance parallel to the external modulation has a sharp peak at $\nu = 2l + 1/2$, considering a spin degree of freedom. We interpret that the resistance peak is induced by an energy gap in parallel direction to the external modulation. The unidirectional lateral superlattice has a same periodic structure with period 92nm as before experiment[7, 13]. The experimental data for the $g = 0.015\mathrm{meV}$ modulation and the 2D electron density $2.8 \times 10^{15}\mathrm{m}^{-2}$ are plotted in the phase diagram Fig. 7 at the half-filled $l+1$-th LL from $l = 3$ to $l = 6$. The all experimental data belong to the orthogonal phase, which means that an energy gap is parallel to the external modulation. This is also consistent with the experiment. We hope that a similar experiment with a higher mobility sample will give more clear evidence for our results.

It seems difficult to understand how the orthogonal phase is realized contrary to the naive expectation that two striped structures tend to be parallel. To understand the reason, it is convenient to consider the striped state in the momentum space. Since the Fermi surface of the striped state is flat as seen in Fig. 1, a perturbation with a wave number vector perpendicular to the Fermi surface affects the total energy most strongly. Therefore the orthogonal phase could be realized in a small external modulation. The point of our theory is that the mean field theory has the flat Fermi surface. The fluctuation around the mean field has been studied but discussions seem unsettled yet[14, 15, 16]. Higher order corrections are expected to be small because the Fermi velocity diverges due to the Coulomb interaction[12]. We estimate $\Delta E_l^{(2)}$ in the RPA approximation to the density correlation function as $\pi_{00}^{\rm RPA}(k) = \pi_{00}(k)/(1 - \tilde{V}(k)\pi_{00}(k))$ where $\tilde{V}(k)$ is the Fourier transformed form of $V(\mathbf{r})$. The results are shown in Fig. 3 by dashed lines. As seen in this figure the correction is small actually.

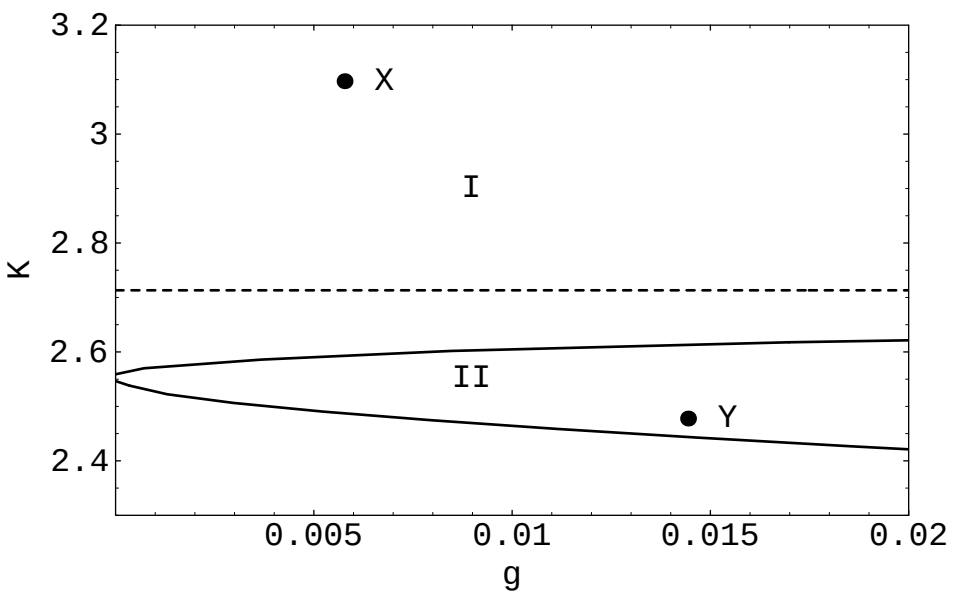

**Figure 6.** The two points $X$ and $Y$ in the phase diagram stand for the experimental data[7, 13] at half-filled third LL.

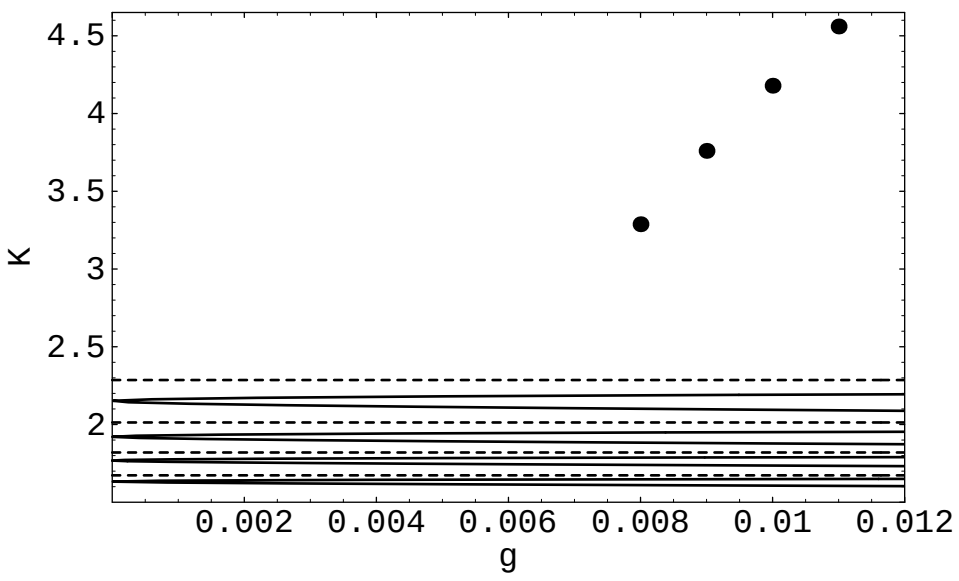

**Figure 7.** The four points stand for the experimental data[7, 13]: $l = 3, 4, 5, 6$ from bottom to top. The four curves stand for the phase boundary near the commensurate wave number for each LL:$l = 6, 5, 4, 3$ from bottom to top.

In summary it is shown that a weak external modulation determines the orientation of the striped state and there are two phases in the 2D parameter space of the strength and wave number of the external modulation, that is, the orthogonal phase and parallel phase. In the former phase, the optimal orientation of the striped state is orthogonal to the external modulation. In the latter phase, the optimal orientation is parallel to the external modulation. The phase diagram is obtained numerically at the half-filled $l + 1$-th ($l = 1, \cdots, 8$) LL. We believe that our findings shed a new light on the origin of an orientation of striped states in quantum Hall systems.

We thank A. Endo and Y. Iye for useful discussions. This work was partially supported by the special Grant-in-Aid for Promotion of Education and Science in Hokkaido University provided by the Ministry of Education,Science, Sports, and Culture, Japan, and by the Grant-in-Aid for Scientific Research on Priority area of Research (B) (Dynamics of Superstrings and Field Theories, Grant No. 13135201) from the Ministry of Education, and by Nukazawa Science Foundation.